\documentclass[final,1p,times]{elsarticle} 

\usepackage{graphicx}
\usepackage{amssymb} 
\usepackage{amsmath}
\usepackage{amsthm} 
\usepackage{lineno}
\usepackage{caption}

\everymath{\displaystyle}

\journal{Nuclear Physics A} 

\begin{document}

\begin{frontmatter} 

\title{Isolation of Flow and Nonflow \\by Two- and Multi-Particle Cumulant Measurements of $v_n$ \\ in $\sqrt{s_{NN}} = 200$ GeV Au+Au Collisions by STAR}

\author{Li Yi (for the STAR\fnref{col1} Collaboration)}
\fntext[col1] {A list of members of the STAR Collaboration and acknowledgements can be found at the end of this issue.}
\address{Purdue University, West Lafayette, U.S.}


\begin{abstract} 

We apply a data-driven method to STAR Au+Au collisions at $\sqrt{s_{NN}}=200$ GeV to isolate $\Delta\eta$-dependent and $\Delta\eta$-independent correlations by using two- and four-particle Q-cumulant $v_{n}$ measurements. The $\Delta\eta$-independent part, dominated by flow, is found to be $\eta$-independent within the STAR TPC of $\pm1$ unit of pseudo-rapidity. The $\Delta\eta$-dependent part may be associated to nonflow, and is responsible for the $\Delta\eta$ drop in the measured two-particle $v_{2}$ cumulant. We combine our result to four- and six-particle cumulants to gain further insights on the nature of flow fluctuations.

\end{abstract} 

\end{frontmatter} 


\section{Introduction}

Heavy ion collisions at the  Relativistic Heavy Ion Collider (RHIC) provide a means to study the Quark Gluon Plasma (QGP). In a non-central collision, the transverse overlap region is anisotropic. The energy density gradient converts the initial coordinate-space anisotropy into the final momentum-space one, generally called flow. As the system expands, the coordinate-space anisotropy diminishes. Hence, flow is sensitive mostly to the early stage of the collision \cite{v2}. Through measurements of anisotropic flow and comparison to hydrodynamic calculations, properties of the early stage of the collision system may be extracted. For example, the shear viscosity to entropy density ratio ($\eta/s$) of the QGP was found to be not much larger than the conjectured quantum limit of $1/4 \pi$ \cite{Kovtun}.

The momentum-space anisotropic flow can be characterized by the Fourier coefficients of the particle azimuthal distribution relative  to a symmetry plane (participant plane) \cite{Art}. However, the participant plane is not experimentally accessible. As a proxy, the event plane is constructed from particle momentum. Similarly, two- and multi-particle correlations are also used to quantify anisotropy. These various flow measurements are therefore contaminated by intrinsic particle correlations unrelated to the participant plane, generally called nonflow \cite{Bor}. Although nonflow at low transverse momentum ($p_{T}$) may be small, it was found that $\eta/s$ is very sensitive to flow \cite{song}. Therefore, it is important to separate the nonflow contributions from flow measurements.


\section{Measurement Methods}

This analysis  attempts to separate flow and nonflow in a data-driven manner \cite{Xu}. The method utilizes two-particle cumulant with one particle at $\eta_{\alpha}$ and another at $\eta_{\beta}$ , and four-particle cumulant with two at $\eta_{\alpha}$ and two at $\eta_{\beta}$. We have
\begin{align}
V \lbrace 2 \rbrace & \equiv \langle\langle 2 \rangle\rangle \equiv v (\eta_{\alpha}) v (\eta_{\beta}) + \sigma (\eta_{\alpha}) \sigma (\eta_{\beta}) + \sigma ' (\Delta\eta) + \delta (\Delta \eta), \label{EqV2} \\
V \lbrace 4 \rbrace & \equiv \sqrt{2 \langle \langle 2 \rangle \rangle^{2} - \langle \langle 4 \rangle \rangle} \approx v (\eta_{\alpha}) v (\eta_{\beta}) - \sigma (\eta_{\alpha}) \sigma (\eta_{\beta}) - \sigma ' (\Delta\eta), \label{EqV4}
\end{align}
where $\Delta\eta=\eta_{\beta}-\eta_{\alpha}$, and we require $\eta_{\alpha}<\eta_{\beta}<0 $. The average flow ($v$) is only a function of $\eta$. $\sigma$ and $\sigma'$ are the flow fluctuation, which may have $\Delta\eta$-dependence component. Nonflow ($\delta$) is generally a function of $\Delta\eta$.

We take the difference between cumulants $V_{n} \lbrace 2 \rbrace$ at $(\eta_{\alpha}, \eta_{\beta})$ and $(\eta_{\alpha}, -\eta_{\beta})$. We do the same for $V_{n} \lbrace 4 \rbrace $. For symmetric collision systems, the $\Delta\eta$-independent terms in Eq.~\eqref{EqV2} and ~\eqref{EqV4} cancel out, leading to Eq.~\eqref{EqDV2} and~\eqref{EqDV4}:
\begin{align}
\Delta V \lbrace 2 \rbrace & \equiv V \lbrace 2 \rbrace (\eta_{\alpha},\eta_{\beta})- V \lbrace 2 \rbrace (\eta_{\alpha},-\eta_{\beta}) 
 \equiv V \lbrace 2 \rbrace (\Delta\eta_{1})- V \lbrace 2 \rbrace (\Delta\eta_{2}) = \Delta \sigma '  + \Delta \delta, \label{EqDV2} \\
\Delta V \lbrace 4 \rbrace & \equiv V \lbrace 4 \rbrace (\eta_{\alpha},\eta_{\beta})- V \lbrace 4 \rbrace (\eta_{\alpha},-\eta_{\beta})  
 \equiv V \lbrace 4 \rbrace (\Delta\eta_{1})- V \lbrace 4 \rbrace (\Delta\eta_{2}) \approx -\Delta \sigma ', \label{EqDV4}
\end{align}
where $
\Delta\sigma'=\sigma'(\Delta\eta_{1})-\sigma'(\Delta\eta_{2}), \Delta\delta = \delta(\Delta\eta_{1})-\delta(\Delta\eta_{2}) , \ $ and $
\Delta\eta_{1}  \equiv \eta_{\beta} - \eta_{\alpha}  , \ 
\Delta\eta_{2} \equiv -\eta_{\beta} - \eta_{\alpha} \label{EqDh}.
$
Assuming the same flow fluctuations $\sigma'$ in both $V\lbrace 2 \rbrace$ and $V\lbrace 4 \rbrace$, we can extract information of nonflow ($\delta$) and $\Delta\eta$-dependent flow fluctuation ($\sigma'$). Using Eq.~\eqref{EqV2} and~\eqref{EqV4}, we can further obtain $v$ and $\sigma$.

\section{Data Analysis and Results}

We analyze the data in Au+Au collisions at $\sqrt{s_{NN}} = 200$ GeV from year 2004 measured by the STAR Time Projection Chamber (TPC) within $|\eta|<1.0$. The Q-cumulant method \cite{Ante, LY} is used with the acceptance correction in calculation. Azimuthal moments are weighted by unit (not pair/quadrant multiplicities).

\begin{figure}[htb]
\begin{center}
\includegraphics[width=0.83\textwidth]{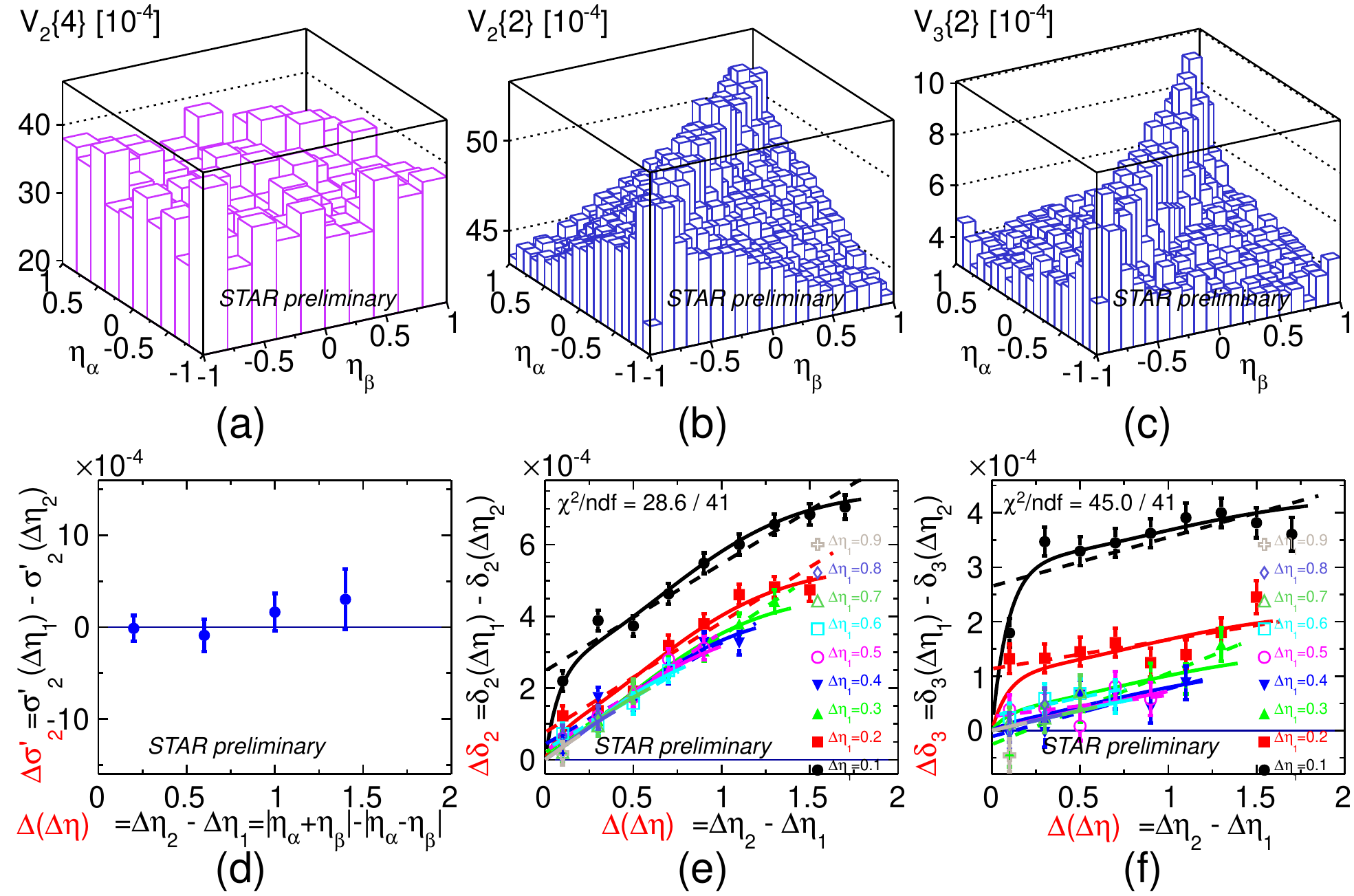}
\end{center}
\caption{(a) The second harmonic four-particle cumulant. (b,c) The second and third harmonic two-particle cumulants. (d) Difference in $V_{2} \lbrace 4 \rbrace $ between $(\eta_{\alpha},\eta_{\beta})$ and $(\eta_{\alpha},-\eta_{\beta})$ as a function of $\Delta\eta_{2} - \Delta\eta_{1} = \lvert 2\eta_{\beta} \rvert$. (e,f) Difference in $V_{2} \lbrace 2 \rbrace $ and $V_{3} \lbrace 2 \rbrace $ for given $\Delta\eta_{1}$ values, respectively. Data are from the 20-30\% centrality of Au+Au collision at $\sqrt{s_{NN}}=200$ GeV.}
\label{legoAndfit}
\end{figure}

Figure~\ref{legoAndfit} (a), (b) and (c) show the cumulant results in Eq.~\eqref{EqV2} and~\eqref{EqV4} as a function of $\eta_{\alpha}$, $\eta_{\beta}$, and (d), (e) and (f) show the quantities in Eq.~\eqref{EqDV2} and~\eqref{EqDV4} for 20-30\% centrality. $\Delta V_{2} \lbrace 4 \rbrace$ is shown in Fig.~\ref{legoAndfit} (d), and is consistent with zero, suggesting that elliptic flow fluctuations are independent of $\Delta\eta$ within the TPC acceptance. This is used in subsequent analysis of data. We further assume the triangular flow fluctuations are also independent of $\Delta\eta$ within the TPC acceptance. With this taken into account, $\Delta V \lbrace 2 \rbrace$ will only have nonflow component. $\Delta V_{2} \lbrace 2 \rbrace$ and $\Delta V_{3} \lbrace 2 \rbrace$ is shown in Fig.~\ref{legoAndfit} (e) and (f), respectively. 

\begin{figure}[htb]
\begin{center}
\includegraphics[width=0.83\textwidth]{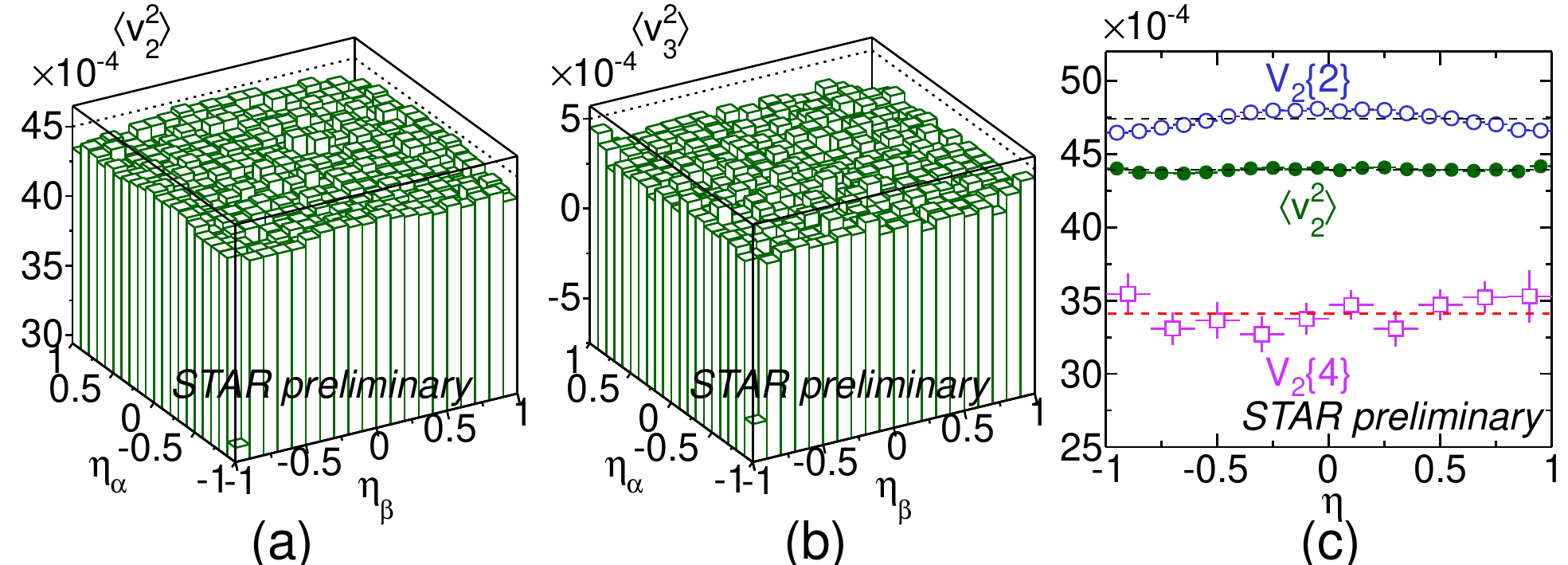}
\end{center}
\caption{(a) $\langle v_{2}^{2} \rangle$. (b) $\langle v_{3}^{2} \rangle$ as a function of $(\eta_{\alpha},\eta_{\beta})$. The parameterized nonflow has been subtracted. (c) the projection of $\eta$ dependence of the decomposed flow $\langle v_{2}^{2} \rangle$ and the raw two- and four-particle cumulants of $v_{2}$ for the 20-30\% centrality of Au+Au collision at $\sqrt{s_{NN}}=200$ GeV.}
\label{figv}
\end{figure}

As the dashed lines shown in Fig.~\ref{legoAndfit} (e) and (f), the nonflow difference between $\Delta\eta_{1}$ and $\Delta\eta_{2}$ appears linear in $\Delta\eta_{2}-\Delta\eta_{1}$, with a slope that is independent of $\Delta\eta_{1}$. The intercept of the linear dependence decreases quickly with $\Delta\eta_{1}$, suggesting a short-range exponential component in nonflow. In other words, the nonflow difference can be described by
\begin{equation}
\displaystyle \delta(\Delta\eta_{1})-\delta(\Delta\eta_{2}) = a [e^{\frac{-\Delta\eta_{1}}{b}}-e^{\frac{-\Delta\eta_{2}}{b}}]+A[e^{\frac{-\Delta\eta_{1}^{2}}{2 \sigma^{2} }} - e^{\frac{-\Delta\eta_{2}^{2}}{2 \sigma^{2}}}]. \label{EqDd}
\end{equation}
A single fit of Eq.~\eqref{EqDd}  with four free parameters $a, b, A, \sigma$ is applied to all the data points, yielding the solid curves in Fig.~\ref{legoAndfit} (e) and (f). The fit results give the following nonflow parameterization:
\begin{equation}
\displaystyle \delta(\Delta\eta) = a e^{\frac{-\Delta\eta}{b}} + A e^{\frac{-\Delta\eta^{2}}{2 \sigma^{2}}}. \label{eqd}
\end{equation}
Subtracting Eq.~\eqref{eqd} from the raw two-particle cumulant 
gives the decomposed `flow' results in Fig.~\ref{figv} (a) and (b). Here the decomposed flow includes the flow fluctuation effect $\langle v(\eta_{\alpha}) v(\eta_{\beta}) \rangle = v(\eta_{\alpha}) v(\eta_{\beta}) + \sigma(\eta_{\alpha}) \sigma(\eta_{\beta})$.

\begin{figure}[htb]
\centering
\begin{minipage}[t]{0.48\textwidth}
\centering
\includegraphics[width=0.65\textwidth]{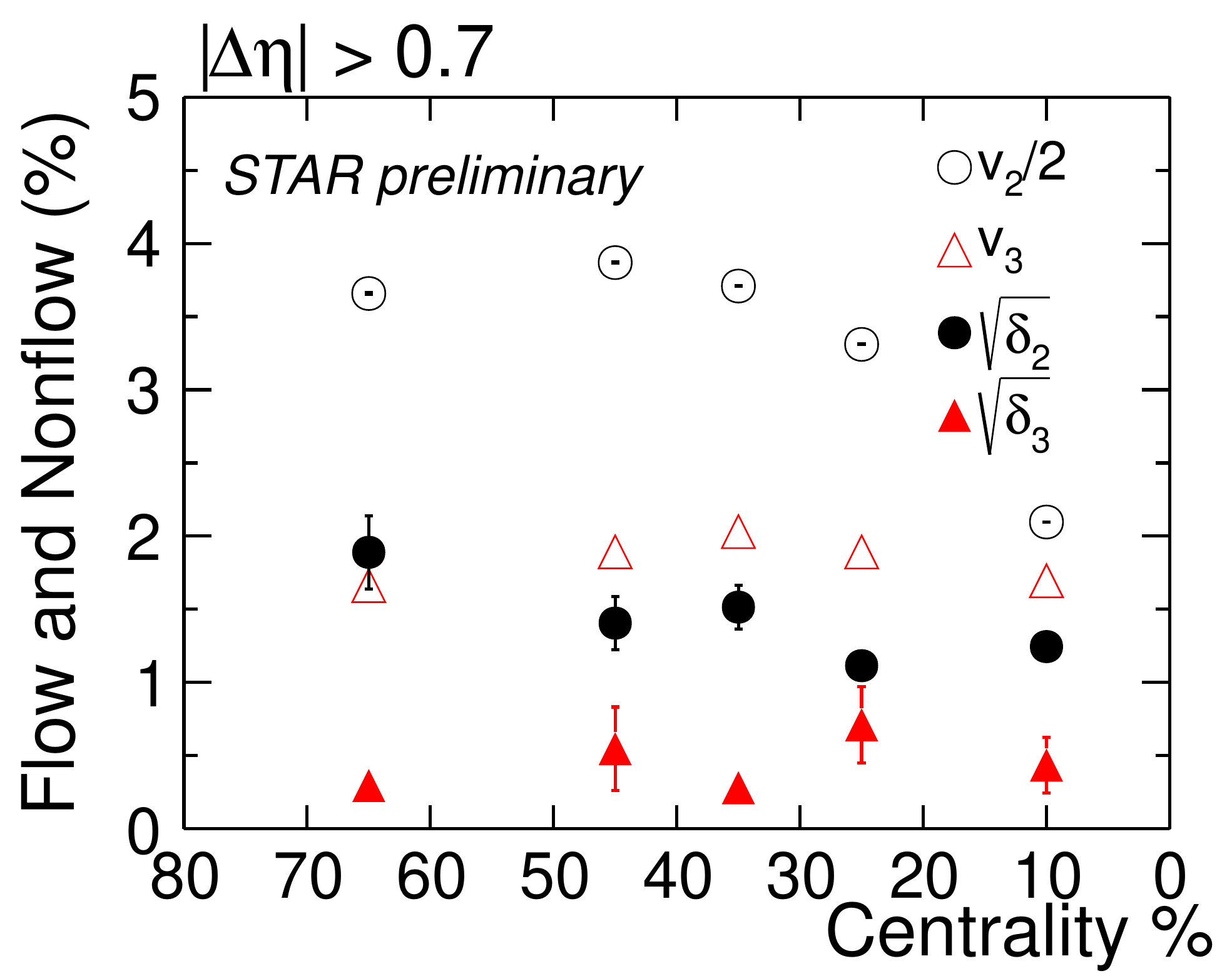}
\caption{the decomposed flow and nonflow with $\eta$-gap $> 0.7$ as a function of centralities}
\label{vdfitcent}
\end{minipage}%
\hspace{4mm}
\begin{minipage}[t]{0.48\textwidth}
\centering
\includegraphics[width=0.65\textwidth]{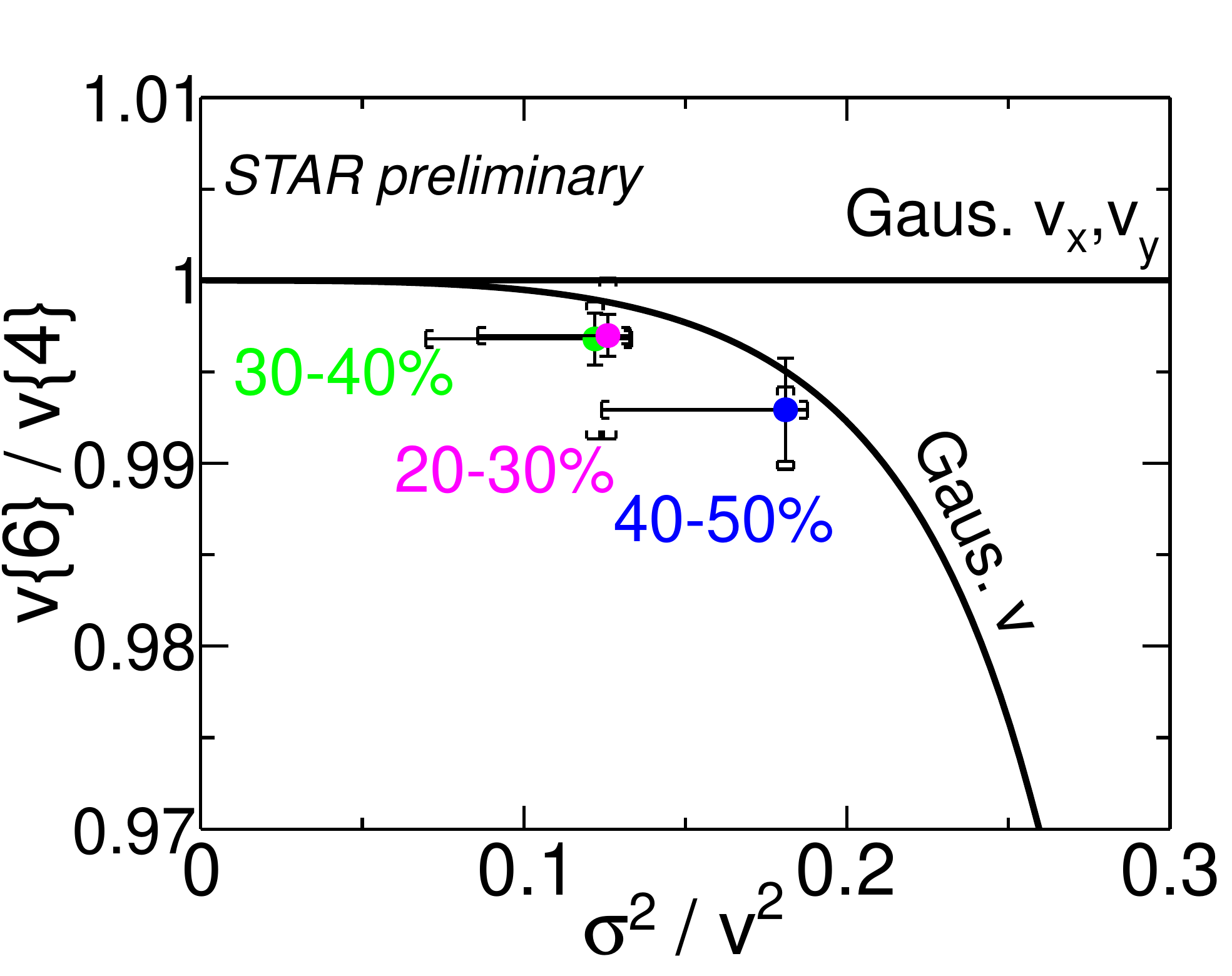}
\caption{$v_{2}\lbrace 6 \rbrace / v_{2}\lbrace 4 \rbrace$ and $\sigma_{2}^{2} /  v_{2}^{2} $. The curves are model predictions of ref.~\cite{Voloshin} and~\cite{LY}}
\label{figv46}
\end{minipage}
\end{figure}

Figure~\ref{figv} (a) and (b) shows that $v v+\sigma \sigma$ is independent of $\eta$ for both $v_{2}$ and $v_{3}$, and Fig.~\ref{legoAndfit} (a) shows $v_{2}v_{2}- \sigma_{2}\sigma_{2}$ is independent of $\eta$. This suggests that $v_{2}$ and its fluctuation are independent of $\eta$. The projections of Fig.~\ref{figv} (a), Fig.~\ref{legoAndfit} (a) and (b) are shown in Fig.~\ref{figv} (c). The difference between the raw two-particle cumulant $V_{2}\lbrace 2 \rbrace $ and the decomposed `flow' $\langle v_{2}^{2}\rangle$ is the nonflow effect. We find that $\delta_{2}/v_{2}^{2}$ is about 4\%, and $\sigma_{2}^{2}/v_{2}^{2}$ is about 13\% for the centrality 20-30\%. Figure~\ref{vdfitcent} shows the decomposed `flow' and nonflow as a function of centrality in Au+Au $\sqrt{s_{NN}}=$ 200 GeV.

The flow fluctuation may also be assessed by four- and six-particle cumulants. Assuming a Gaussian fluctuation of the flow magnitude, we have $v \lbrace 6 \rbrace / v \lbrace 4 \rbrace \approx 1 - \sigma^{6} / 3v^{6} \text{ , when } \ \sigma^{2} / v^{2} \ll 1 $ \cite{LY}. 
On the other hand, independent Gaussian fluctuations of $v_{x}, v_{y}$ would give $v \lbrace 6 \rbrace = v \lbrace 4 \rbrace$ \cite{Voloshin}. We show in Fig.~\ref{figv46} $v_{2}\lbrace6\rbrace/v_{2}\lbrace4\rbrace$ vs the fluctuation results from our decomposition method. The systematic errors are estimated by the different moment weight methods and non-uniform acceptance correction. The $v_{2}$ results, while favoring Gaussian fluctuations in the flow magnitude, may be also consistent with Gaussian fluctuations in the individual flow components. A firmer conclusion on the nature of flow fluctuations await for more statistics.  


\section{Summary}
We
analyze two- and four-particle Q-cumulants between pseudo-rapidity bins in Au+Au collisions at $200$ GeV from STAR.
Exploiting the collision symmetry about mid-rapidity, we isolate the $\Delta\eta$-dependent
and $\Delta\eta$-independent correlations in the data. Elliptic flow fluctuations appear independent of $\Delta\eta$ within the TPC acceptance, as deduced from the four-particle cumulant measurement. The method does not make assumptions about the $\eta$ 
dependence of flow. Our isolated $\Delta\eta$-independent part of our data is dominated by
flow, with a relatively small away-side nonflow, and is found to be $\eta$-independent within the STAR TPC of $\pm$ 1 unit of
pseudo-rapidity. The $\Delta\eta$-dependent part is nonflow and is found to decrease with $\Delta\eta$. In the 20-30\% Au+Au collisions at $\lvert \eta \rvert <1 $, and $p_{T} < 2$ GeV/$c$, the elliptic nonflow relative to flow $\delta_{2}/\langle v_{2}^{2} \rangle$ is found to be $4\%$ and the relative fluctuation $\sigma_{2}^{2}/ \langle v_{2}^{2} \rangle$ is 13\%. The flow fluctuation in conjunction with the 6- and 4-particle cumulants flow ratio seems to favor Gaussian fluctuations in the flow magnitude.



\begin{thebibliography}{00} 
\bibitem{v2} J. Y. Ollitrault, Phys. Rev. D 46, 229 (1992)
\bibitem{Kovtun} P. K. Kovtun, D. T. Son, and A. O. Starinets, Phys. Rev. Lett. 94, 111601 (2005)
\bibitem{Art} A. M. Poskanzer and S. A. Voloshin, Phys. Rev. C 58, 1671 (1998)
\bibitem{Bor} N. Borghini, P. M. Dinh, and J. Y. Ollitrault, Phys. Rev. C 62, 034902 (2000)
\bibitem{song} H. Song {\itshape et al\/}, Phys. Rev. Lett. 106, 192301 (2011)
\bibitem{Xu} L. Xu {\itshape et al\/}, Phys. Rev. C 86, 024910 (2012)
\bibitem{Ante} A. Bilandzic, R. Snellings, and  S. Voloshin, Phys. Rev. C 83, 044913 (2011)
\bibitem{LY} L. Yi, F. Wang, and A. Tang, arXiv: 1101.4646 $[$nucl-ex$]$ 
\bibitem{Voloshin} S. A. Voloshin, A. M. Poskanzer, A. Tang, and G. Wang, Phys. Lett. B 659, 537 (2008) 



\end{thebibliography}
\end{document}